\documentclass[conference]{IEEEtran}
\IEEEoverridecommandlockouts
\usepackage{cite}
\usepackage{amsmath,amssymb,amsfonts}
\usepackage{algorithmic}
\usepackage{graphicx}
\usepackage{textcomp}
\usepackage{xcolor}
\def\BibTeX{{\rm B\kern-.05em{\sc i\kern-.025em b}\kern-.08em
    T\kern-.1667em\lower.7ex\hbox{E}\kern-.125emX}}

\usepackage[numbers]{natbib}
\usepackage{mathtools}
\usepackage{tabularx}
\usepackage{subcaption}
\usepackage{tikz}
\usetikzlibrary{shapes}
\usepackage{balance}

\begin{document}

\title{Spatial-Temporal Cluster Relations: A Foundation for Trajectory Cluster Lifetime Analysis\\
}

\author{
\IEEEauthorblockN{Ivens Portugal}
\IEEEauthorblockA{
\textit{University of Waterloo}\\
Waterloo, Canada \\
iportugal@uwaterloo.ca}
\and
\IEEEauthorblockN{Paulo Alencar}
\IEEEauthorblockA{
\textit{University of Waterloo}\\
Waterloo, Canada \\
palencar@cs.uwaterloo.ca}
\and
\IEEEauthorblockN{Donald Cowan}
\IEEEauthorblockA{
\textit{University of Waterloo}\\
Waterloo, Canada \\
dcowan@csg.uwaterloo.ca}
}

\maketitle

\begin{abstract}
Spatial-temporal data, that is information about objects that exist at a particular location and time period, are rich in value and, as a consequence, the target of so many initiative efforts. Clustering approaches aim at grouping datapoints based on similar properties for classification tasks. These approaches have been widely used in domains such as human mobility, ecology, health and astronomy. However, clustering approaches typically address only the static nature of a cluster, and do not take into consideration its dynamic aspects. A desirable approach needs to investigate relations between dynamic clusters and their elements that can be used to derive new insights about what happened to the clusters during their lifetimes. A fundamental step towards this goal is to provide a formal definition of spatial-temporal cluster relations. This report introduces, describes, and formalizes 14 novel spatial-temporal cluster relations that may occur during the existence of a cluster and involve both trajectory-cluster membership conditions and cluster-cluster comparisons. We evaluate the proposed relations with a discussion on how they are able to interpret complex cases that are difficult to be distinguished without a formal relation specification. We conclude the report by summarizing our results and describing avenues for further research.
\end{abstract}


\section{Introduction}

Spatial-temporal data refers to pieces of information in which their spatial and the temporal aspects are relevant \cite{Wang2000202, Kent2002213}. Some examples are a person walking on the street or a taxi driving on a busy road. Since location tracking devices became popular, spatial-temporal data is highly available, and their analysis is increasingly important.

A well-known, widely used spatial-temporal data analysis technique is clustering \cite{Tan2018}. Clusters are groups of similar elements, such that elements in a cluster are similar to each other and dissimilar to those outside the cluster \cite{Tan2018}. These approaches have been widely used in domains such as human mobility, ecology, health and astronomy. Two popular clustering approaches are the distance-based K-Means \cite{Lloyd1982129} and the density-based DBSCAN \cite{Ester1996}. These approaches are able to identify static clusters (i.e. clusters that do not move) that can be useful for classification of new data or outlier detection. Dynamic, moving, or spatial-temporal clusters are clusters that move on the space throughout time \cite{Kisilevich2010, Ansari2019}. Their analysis is significantly more difficult because of the many movement patterns that clusters may follow. Some popular approaches are ST-DBSCAN \cite{Birant2007208} and the work in \cite{Chen20152575}.

Despite recent advances, clustering approaches typically address only the static nature of a cluster, and do not take into consideration its dynamic aspects. Specifically, during traditional analysis, spatial-temporal clustering approaches focus on identifying the clusters, but do not capture the dynamic cluster-cluster and cluster-trajectory relations that can be used to derive new insights about what happened to the clusters during their lifetimes. A fundamental required step in this direction is to provide a formal definition of spatial-temporal cluster relations.

This report introduces, describes, and formalizes 14 cluster relations that can be used to identify higher level characteristics and behaviors of clusters and involve both trajectory-cluster membership conditions and cluster-cluster comparisons. These new relations can lead to new insights in cluster analysis. For example, dynamic cluster relations such as merge and split can provide valuable information to support decision making involving situations in which clusters (e.g., taxis, animals) grow, shrink, merge, or split.

This report is divided into the following sections. Section \ref{relatedwork} details some background information about spatial-temporal data, spatial and temporal relations, as well as gives insights about how the 14 relations were identified. Section \ref{main} introduces, describes, and formalizes the 14 relations. Section \ref{examples} discusses how these relations are able to interpret complex cases that are difficult to be distinguished without a formal relation specification. Finally Section \ref{conclusions} concludes the report.

\section{Related Work}
\label{relatedwork}

\subsection{Spatial-Temporal Data}

Spatial data is information about an element, such that its location in a space is considered \cite{Forlizzi1998332, Abdul-Rahman20081}. A common example is geographical spatial data, where objects exist somewhere on the planet. For example, the location of a store is spatial data. Other types of spatial data are molecules and their place in the human body, planets on the solar system or the universe, or even hypothetical multidimensional objects in a multidimensional space. Although the most common way to refer to spatial data is in terms of its latitude and longitude (sometimes even altitude), this is not the only one. Coordinates and relative positioning (radius in meters from a centroid) are also used. Usually, spatial data is reduced to a zero dimensional point for calculations and comparisons (e.g. centroid). However, when relevant, one (line), two (region), or even more (cube, hypercube) dimensions are considered in the calculations.

Temporal data is information about an element, such that the time of its existence is considered \cite{Jensen199936, Laxman2006173}. Usually, temporal data describe events. For example, an exam started at 10am and finished at 11am. Temporal data is usually instantaneous, but it can also be described in terms of a duration. Some concerns when describing temporal data are time granularity and time zones. Time granularity refers to the richness of the description of time. For example, it is sufficient to describe a holiday by the day it happens, but the description of bank transactions should include hours, minutes, seconds, and perhaps milliseconds. Time zones are regions of the planet where a common standard time is used. The Americas, Europe, and Asia are in different time zones, and events described in one of these regions should be ``temporally translated'' to another region before any analysis. To avoid confusion, a Universal Time Coordinate is used.

\begin{figure}
  \includegraphics[width=\linewidth]{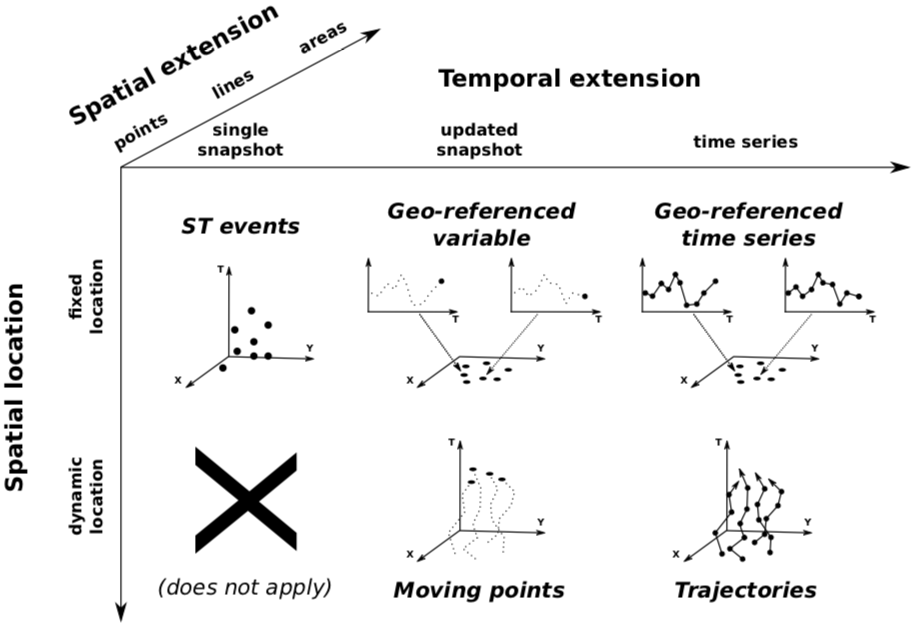}
  \caption{Classification of spatial-temporal data based on its dimensions and shape characteristics. \cite{Kisilevich2010}}
  \label{stclustering}
\end{figure}

Spatial-temporal data refers to elements whose spatial and temporal dimensions are considered. A common example is a vehicle on a road, such that their location change with time. Since this type of data is described based on its spatial and temporal characteristics, concerns when describing these dimensions are relevant here. Although it is intuitive to imagine a spatial-temporal object moving on space as time passes, this is not a strong requirement. For example, a sports match having its location and start and end times is an example of spatial data. Another example is weather balloons, where information about the weather is captured at several different locations at different timestamps. In fact, the work in \cite{Kisilevich2010} classifies spatial-temporal data based on the relevance of its spatial and temporal dimensions, and its shape. See Figure \ref{stclustering}. Note the difference between moving points and trajectories. Both refer to spatial-temporal objects that move throughout time, but only the most recent snapshot of information can be accessed from the former, whereas the entire history of movement is available from the latter.

\subsection{Spatial-Temporal Relations}

Spatial relations describe how spatial objects interact in the space. There are three major types described in the literature: topological, directional, and distance. Topological relations specify relations of containment, coverage, or intersection. For example, a house \emph{contained} in a city, or the path left by two vehicles \emph{crosses}. Note that topological relations can be described for objects of any number dimensions. The Dimensionally Extended nine-Intersection Model (DE-9IM) \cite{Clementini1993277} expresses topological relations in a systematic way for objects in a two-dimensional space. It describes objects in terms of three of their regions: interior (I), boundary (B), and exterior (E). Then a matrix is constructed (See Figure \ref{de9im}) and the dimension of the intersection of these regions in both objects is evaluated. As an example, the \textit{contains} relation is defined by the matrix in Figure \ref{contains} and the matrix for \textit{equals} is found in Figure \ref{equals}. The set of spatial relations that can be derived from the DE-9IM model include the following 10 relations: \textit{equals}, \textit{disjoint}, \textit{touches}, \textit{contains}, \textit{covers}, \textit{intersects}, \textit{within}, \textit{coveredBy}, \textit{crosses}, and \textit{overlaps},

\begin{figure*}
\centering
\[
\text{DE9IM}(a,b) = 
\begin{bmatrix}
\text{dim}(I(a)\cap I(b)) = 2 & \text{dim}(I(a)\cap B(b)) = 1 & \text{dim}(I(a)\cap E(b)) = 2 \\
\text{dim}(B(a)\cap I(b)) = 1 & \text{dim}(B(a)\cap B(b)) = 0 & \text{dim}(B(a)\cap E(b)) = 1 \\
\text{dim}(E(a)\cap I(b)) = 2 & \text{dim}(E(a)\cap B(b)) = 1 & \text{dim}(E(a)\cap E(b)) = 2
\end{bmatrix}
\]
\caption{The DE-9IM model matrix for spatial relation specification.}
\label{de9im}
\end{figure*}

\begin{figure}
\centering
\begin{subfigure}{0.45\linewidth}
\[
\begin{bmatrix}
T & * & * \\
* & * & * \\
F & F & *
\end{bmatrix}
\]
\caption{The \textit{contains} relation.}
\label{contains}
\end{subfigure}
\begin{subfigure}{0.45\linewidth}
\[
\begin{bmatrix}
T & * & F \\
* & * & F \\
F & F & *
\end{bmatrix}
\]
\caption{The \textit{equals} relation.}
\label{equals}
\end{subfigure}
\caption{The \textit{contains} and \textit{equals} relations derived from the DE-9IM model.}
\label{cede9im}
\end{figure}

Directional spatial relations describe the relative positioning between spatial objects. For example, the pharmacy is \emph{on the left of} the restaurant. They are mainly divided into two groups: the internal and external relations. As the name implies, internal directional spatial relations describe the positioning of an object that is contained in another object. Notice that higher dimensions are necessary. Some examples of internal relations are \textit{left}, \textit{on the back}, and \textit{athwart}. Conversely, external directional spatial relations describe the positioning of objects without a containment relation. Some examples are \textit{on the right of}, \textit{behind}, and \textit{in front of}.

Distance spatial relations take into consideration the proximity between objects when describing them. One example is the train is \emph{near} the road. Notice that, the concept of near, far, and other distance relations are relative to the context. The moon is relatively close to the Earth when compared to the Sun, but two vehicles at the Earth-Moon distance are very far. Some distance relations are \textit{at}, \textit{nearby}, and \textit{far away}.

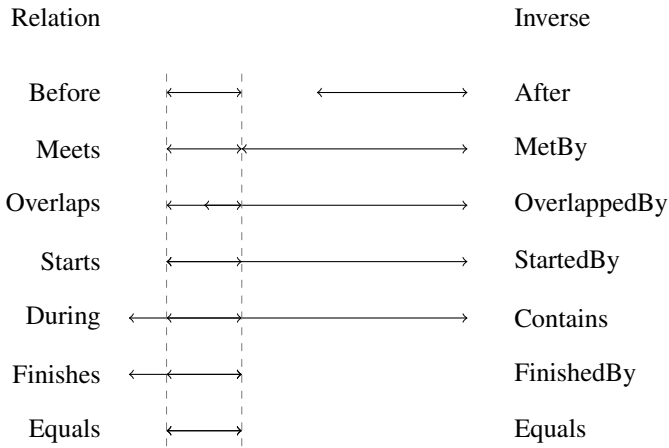
\begin{figure}
\centering
\begin{tikzpicture}

\node [anchor=east] at (-1.75, 5.5) {Relation};
\node [anchor=west] at ( 3.5, 5.5) {Inverse};

\draw [<->] (  -1,4.5) -- (0,4.5);
\draw [<->] (   1,4.5) -- (3,4.5);
\node [anchor=east] at (-1.75, 4.5) {Before};
\node [anchor=west] at ( 3.5, 4.5) {After};

\draw [<->] (  -1,3.75) -- (0,3.75);
\draw [<->] (   0,3.75) -- (3,3.75);
\node [anchor=east] at (-1.75, 3.75) {Meets};
\node [anchor=west] at ( 3.5, 3.75) {MetBy};

\draw [<->] (  -1,3) -- (0,3);
\draw [<->] (-0.5,3) -- (3,3);
\node [anchor=east] at (-1.75, 3) {Overlaps};
\node [anchor=west] at ( 3.5, 3) {OverlappedBy};

\draw [<->] (  -1,2.25) -- (0,2.25);
\draw [<->] (  -1,2.25) -- (3,2.25);
\node [anchor=east] at (-1.75, 2.25) {Starts};
\node [anchor=west] at ( 3.5, 2.25) {StartedBy};

\draw [<->] (-1  ,1.5) -- (0,1.5);
\draw [<->] (-1.5,1.5) -- (3,1.5);
\node [anchor=east] at (-1.75, 1.5) {During};
\node [anchor=west] at ( 3.5, 1.5) {Contains};

\draw [<->] (-1  ,0.75) -- (0,0.75);
\draw [<->] (-1.5,0.75) -- (0,0.75);
\node [anchor=east] at (-1.75, 0.75) {Finishes};
\node [anchor=west] at ( 3.5, 0.75) {FinishedBy};

\draw [<->] (  -1,0) -- (0,0);
\draw [<->] (  -1,0) -- (0,0);
\node [anchor=east] at (-1.75, 0) {Equals};
\node [anchor=west] at ( 3.5, 0) {Equals};

\draw [dashed, gray] (-1,4.75) -- (-1,-0.25);
\draw [dashed, gray] ( 0,4.75) -- ( 0,-0.25);

\end{tikzpicture}
\caption[captionfortemporal]{A visual representation of the elementary temporal relations. Adapted from W3C's temporal ontology specification\protect\footnotemark}
\label{tr-figure}
\end{figure}

Temporal relations describe the occurrence of two events relative to each other. For example, the meeting happened \emph{before} the lunch. The description of temporal relations can be done between time periods and timestamps. Notice that they are very similar to spatial relations between lines and points in a one-dimensional space, because of their nature. The World Wide Web Consortium (W3C)\footnote{https://www.w3.org} has an entire document dedicated to the description and formalization of a temporal ontology\footnote{https://www.w3.org/TR/owl-time}. The document describes seven topological temporal relations between time periods. They are described in Table \ref{tr-table} and a visual representation is depicted in Figure \ref{tr-figure}.

\begin{table}
\caption{A description of topological temporal relations.}
\label{tr-table}
\begin{tabularx}{\linewidth}{l l X}
Relation & Inverse      & Description                                                                         \\
\hline
Before   & After        & One time period occurs subsequent to another, but not immediately.                  \\
Meets    & MetBy        & One time period occurs subsequent to another, without any delay.                    \\
Overlaps & OverlappedBy & There exists some temporal intersection between the occurrence of two time periods. \\
Starts   & StartedBy    & Two time periods start at the same time.                                            \\
During   & Contains     & One time period starts and ends while the other time period is occurring.           \\
Finishes & FinishedBy   & Two time periods end at the same time.                                              \\
Equals   & Equals       & Two time periods start and end at the same time.                                    \\
\hline
\end{tabularx}
\end{table}

Spatial temporal relations describe the interactions between two spatial-temporal objects in a place across time. The literature can still be enriched with novel ways to systematically define spatial-temporal relations, or with the description of such relations in several different conditions (e.g number of objects, their dimensions, the space dimensions). The work in \cite{Mazimpaka2017} attempts to formalize and describe one such list of spatial-temporal relations. It divides relations based on the geometry and the presence or absence of moment of objects. Among the many relations describe are \textit{approaching}, \textit{arriving}, \textit{encounter}, \textit{meeting}, \textit{separating}, \textit{moving-away}, \textit{leaving}, \textit{passing}, \textit{jointly-moving}, and \textit{separately-moving}.
\footnotetext{https://www.w3.org/TR/owl-time}

This is by no means an extensive list of spatial, temporal, or spatial-temporal relations or their types. In fact, a detailed description and formalization of each of these types, relations, and additional relations, in different dimensions and situations, can be found in \cite{Aiello2007}.

\section{Dynamic Spatial-Temporal Cluster Relations}
\label{main}

In this section, we introduce, describe, and formalize 14 spatial-temporal cluster relations to augment the knowledge about what happens to clusters as they move throughout time. These relations are derived from the list of relations identified in \cite{Portugal20193452} and \cite{Portugal20194534}, in which each relation was further divided in other cases. Table \ref{relations} lists each relation alongside a natural language description. Relations are divided into two groups, existential (the first two relations) and spatial-temporal (the remaining relations), based on their nature. Following sections detail each relation. We end this section with a couple of examples to illustrate the usefulness of these relations.

\begin{table*}
\centering
\caption{The 14 cluster spatial-temporal relations.}
\label{relations}
\begin{tabularx}{\linewidth}{c l X}
\# & Relation & Description \\
\hline
1 & START   & A relation that represents the beginning of a cluster.         \\
2 & END     & A relation that represents the end of a cluster.               \\ \hline
3 & T\_ENTER & A relation that represents a trajectory entering a cluster. \\
4 & T\_LEAVE & A relation that represents a trajectory leaving a cluster. \\
5 & GROUP   & A relation that represents the creation of a cluster, when trajectories come together at a location at the same time. \\
6 & DISPERSE & A relation that represents the end of a cluster, when each trajectories that compose the cluster follow a different path. \\
7 & MERGE    & A relation that represents the creation of a cluster, when two or more clusters come together at a location at the same time. \\
8 & SPLIT    & A relation that represents the end of a cluster, when the cluster is divided into smaller clusters that follow a different path. \\
9 & C\_ENTER  & A relation that represents a cluster entering another cluster. This relation is described from the perspective of the cluster who received another one. \\
10 & C\_LEAVE  & A relation that represents a cluster leaving another cluster. This relation is described from the perspective of the cluster who released another one. \\
11 & JOIN     & A relation that represents a cluster entering another cluster. This relation is similar to C\_ENTER, but described from the perspective of the cluster who entered another one. \\
12 & DETACH   & A relation that represents a cluster leaving another cluster. This relation is similar to C\_LEAVE, but described from the perspective of the cluster who left another one. \\
13 & C\_IN     & A relation that represents when two clusters exchange a cluster of trajectories. This relation is described from the perspective of the cluster who released the exchanged cluster. \\
14 & C\_OUT    & A relation that represents when two clusters exchange a cluster of trajectories. This relation is described from the perspective of the cluster who received the exchanged cluster. \\
\hline
\end{tabularx}
\end{table*}

Let $TRAJ = \{traj_1, \ldots, traj_m\}$ be a set of $m$ trajectories. Let $traj_j$ be trajectory $j$. Note that trajectories exist throughout time. In that case, $traj_j = [traj_{j,t}]$ is a vector of trajectories, where $traj_{j,t}$ represents trajectory $j$ during timestamp $t$.

Let $C = \{c_1, \ldots, c_n\}$ be a set of $n$ clusters. Let $c_i$ be cluster $i$. Note that a cluster also exists throughout time. For that reason, $c_i$ is a vector of static clusters $c_{i,t}$, i.e. $c_i = [c_{i,t}]$, where $t$ denotes discrete timestamps. Let $c_{i,t}$ be cluster $i$ during timestamp $t$. Once a timestamp is set, a cluster is defined by the trajectories it contains. For example, in a prototype-based cluster, a cluster is the set of all trajectories within a specified radius, and in a density-based cluster, a cluster is the set of all of its core and border points (where points represent trajectories). Therefore, $c_{i,t} = \{traj_{j,t}\}$ during timestamp $t$, for all $traj_{j,t}$ that satisfies the particular clustering technique being used. Note that, since density-based approaches are relevant in our future work, clusters are not defined based on their centroids. Also note that we assume that a valid cluster has at least $min\_cluster$ trajectories.

\subsubsection{START}
The START relation characterizes the beginning of a cluster. See Figure \ref{start} for a visual representation. This relation is heavily dependent on the clustering approach being used. For example in prototype-based cluster approaches, such as K-MEANS, a cluster $c_i$ undergoes a START relation if
\begin{itemize}
	\item $c_{i,t-1} = \varnothing$
	\item there exists at least one trajectory $traj_{j,t}$ such that $distance(\hat{c}_{i,t}, traj_{j,t}) < distance(\hat{c}_{k,t}, traj_{j,t})$ for $k = 1, \ldots, n$, $k \neq i$,
\end{itemize}
where $\hat{c}_{i,t}$ denotes the centroid of cluster $c_i$ during timestamp $t$. Note that this definition does not consider an empty cluster as a valid cluster.

The first condition simply states that the cluster does not exist in the previous timestamp. The second condition is a requirement for a prototype-based cluster, based on a centroid, to exist, which is that there exists enough trajectories (or data points) close to the centroid and far from other centroids.

In a density-based approach, such as DBSCAN, a cluster $c_i$ undergoes a START relation if
\begin{itemize}
	\item $c_{i,t-1} = \varnothing$
	\item there exists at least $min\_cluster$ trajectories (or data points), such that given any of these trajectories, say $traj_{j,t}$, there exists $minPts$ of these trajectories, each of them individually represented by $traj_{k,t}$, such that $distance(traj_{j,t}, traj_{k,t}) < eps$,
\end{itemize}
where $eps$ is a parameter that denotes the radius of a core point and is used to calculate density.

The first condition simply states that the cluster does not exist in the previous timestamp. The second condition is a requirement for a density-based cluster to exist, which is that there exists enough trajectories (or data points) mutually close, increasing the density.

\subsubsection{END}
The END relation characterizes the end of a cluster. See Figure \ref{end} for a visual representation. This relation is the semantic inverse of the START relation. The END relation is also dependent on the clustering approached being used. For example in prototype-based cluster approaches, such as K-MEANS, a cluster $c_i$ undergoes an END relation if
\begin{itemize}
	\item $c_{i,t-1} \neq \varnothing$
	\item for all trajectories $traj_{j,t}$, such that $traj_{j,t-1} \in c_{i,t-1}$, there exists $k \neq i$, such that $distance(\hat{c}_{i,t}, traj_{j,t}) > distance(\hat{c}_{k,t}, traj_{j,t})$ for some $k = 1, \ldots, n$, $k \neq i$,
\end{itemize}
where $\hat{c}_{i,t}$ denotes the centroid of cluster $c_i$ during timestamp $t$. This definition assumes that an empty cluster is not a valid cluster.

The first condition simply states that the cluster does exist in the previous timestamp. The second condition falsifies the requirement for a prototype-based cluster to exist, by asserting that there does not exist trajectories closer to a centroid than to other centroids.

In a density-based approach, such as DBSCAN, a cluster $c_i$ undergoes a END relation if
\begin{itemize}
	\item $c_{i,t-1} \neq \varnothing$
	\item there does not exist enough ($min\_cluster$) trajectories (or data points) in $c_i$ such that for any such trajectory, say $traj_{j,t}$, there exists $minPts$ trajectories in $c_i$, each of them individually represented by $traj_{k,t}$, such that $distance(traj_{j,t}, traj_{k,t}) < eps$,
\end{itemize}
where $eps$ is a parameter that denotes the radius of a core point and is used to calculate density.

The first condition simply states that the cluster does exist in the previous timestamp. The second condition falsifies the requirement for a density-based cluster to exist, by asserting that trajectories that would form a cluster are not mutually close or do not exist in enough number.

\subsubsection{GROUP}

Intuitively, a cluster $c_i$ undergoes a GROUP relation when a group of enough trajectories decides to move together. See Figure \ref{group} for a visual representation.
Formally, a cluster $c_i$ undergoes a GROUP relation if the following three conditions are satisfied:

\begin{itemize}
	\item $c_{i,t-1} = \varnothing$
	\item $c_{i,t} \neq \varnothing$
	\item there does not exist cluster $c_{j,t-1}$ such that $|c_{i,t} \cap c_{j,t-1}| \ge min\_cluster$.
\end{itemize}

The first and the second conditions simply assert the creation of the cluster. The third condition states that the number of trajectories forming the new cluster and that come from a preexisting cluster is not enough to form a cluster by themselves. If that was the case, the relation would be a DETACH.

\subsubsection{DISPERSE}

Intuitively, a cluster $c_i$ undergoes a DISPERSE relation when its trajectories decide to go to different ways. See Figure \ref{disperse} for a visual representation.
Formally, a cluster $c_i$ undergoes a DISPERSE relation if the following three conditions are satisfied:

\begin{itemize}
	\item $c_{i,t-1} \neq \varnothing$
	\item $c_{i,t} = \varnothing$
	\item there does not exist cluster $c_{j,t}$ such that $|c_{i,t-1} \cap c_{j,t}| \ge min\_cluster$.
\end{itemize}

The first and the second conditions simply assert the ending of a cluster. The third condition states that the number of trajectories spreading from the ending cluster going to an existing cluster should not be enough to form a cluster by themselves. If that was the case, the relation would be a JOIN.

\subsubsection{JOIN}

Intuitively, a cluster $c_i$ undergoes a JOIN relation when it joins an existing cluster. See Figure \ref{join} for a visual representation.
Formally, a cluster $c_i$ undergoes a JOIN relation if
\begin{itemize}
	\item $c_{i,t-1} \neq \varnothing$
	\item $c_{i,t} = \varnothing$
	\item there exists only one cluster $c_{j,t}$ such that $|c_{i,t-1} \cap c_{j,t}| \ge min\_cluster$.
\end{itemize}

The first and the second conditions simply assert the ending of a cluster. The third condition states that there exists one, and only one, cluster where enough trajectories go to. If that was not the case, then the relation would be a DISPERSE. If there existed more than one cluster where trajectories came from, then the relation would be a SPLIT.

\subsubsection{DETACH}

Intuitively, a cluster $c_i$ undergoes a DETACH relation when it leaves an existing cluster. See Figure \ref{detach} for a visual representation.
Formally, a cluster $c_i$ undergoes a DETACH relation if
\begin{itemize}
	\item $c_{i,t-1} = \varnothing$
	\item $c_{i,t} \neq \varnothing$
	\item there exists only one cluster $c_{j,t-1}$ such that $|c_{i,t} \cap c_{j,t-1}| \ge min\_cluster$.
\end{itemize}

The first and the second conditions simply assert the creation of a cluster. The third condition states that there exists one, and only one, cluster where enough are coming from. If that was not the case, then the relation would be a GROUP. If there existed more than one cluster where trajectories went to, then the relation would be a MERGE.

\subsubsection{MERGE}

Intuitively, a cluster $c_i$ undergoes a MERGE relation when it is created by the combination (or union) of two or more clusters. See Figure \ref{merge} for a visual representation.
Formally, a cluster $c_i$ undergoes a MERGE relation if
\begin{itemize}
	\item $c_{i,t-1} = \varnothing$
	\item $c_{i,t} \neq \varnothing$
	\item there exists at least two clusters $c_{j,t-1}$ and $c_{k,t-1}$, $j \neq i$, $k \neq i$, such that $|c_{i,t} \cap c_{j,t-1}| \ge min\_cluster$ and $|c_{i,t} \cap c_{k,t-1}| \ge min\_cluster$.
\end{itemize}

The first and the second conditions simply assert the creation of a cluster. The third condition states that at least two clusters sent enough trajectories to take part in the new cluster formation. If that was not the case, the relation would be either a GROUP or a C\_ENTER.

\subsubsection{SPLIT}

Intuitively, a cluster $c_i$ undergoes a SPLIT relation when it disappears by begin divided into two or more clusters. See Figure \ref{split} for a visual representation.
Formally, a cluster $c_i$ undergoes a SPLIT relation if
\begin{itemize}
	\item $c_{i,t-1} \neq \varnothing$
	\item $c_{i,t} = \varnothing$
	\item there exists at least two clusters $c_{j,t}$ and $c_{k,t}$, $j \neq i$, $k \neq i$, such that $|c_{i,t-1} \cap c_{j,t}| \ge min\_cluster$ and $|c_{i,t-1} \cap c_{k,t}| \ge min\_cluster$.
\end{itemize}

The first and the second conditions simply assert the ending of a cluster. The third condition states that the ending cluster sent enough trajectories left to at least two clusters. If that was not the case, the relation would be either a DISPERSE or a C\_LEAVE.

\subsubsection{C\_ENTER}

Intuitively, a cluster $c_i$ undergoes a C\_ENTER relation when it receives a group of trajectories from a cluster (enough to be a cluster themselves), ending the sender cluster. Note that this relation is similar to the JOIN relation, but it is described from the perspective of the cluster who received another cluster. See Figure \ref{center} for a visual representation.
Formally, a cluster $c_i$ undergoes a C\_ENTER relation if
\begin{itemize}
	\item $c_{i,t-1} \neq \varnothing$
	\item $c_{i,t} \neq \varnothing$
	\item there exists a cluster $c_{j,t-1}$ such that $|c_{i,t} \cap c_{j,t-1}| \ge min\_cluster$
	\item $c_{j,t} = \varnothing$.
\end{itemize}

The first and the second conditions simply assert that the cluster existed on both timestamps. The third condition states that enough trajectories entered the cluster coming from a previously existing cluster. The fourth condition asserts that the previously existing cluster ended. If that was not the case, the relation would be a C\_IN.

\subsubsection{C\_LEAVE}

Intuitively, a cluster $c_i$ undergoes a C\_LEAVE relation when a group of trajectories (enough to form a new cluster) leaves the cluster, creating a new cluster. Note that this relation is similar to the DETACH relation, but it is described from the perspective of the cluster who released another cluster. See Figure \ref{cleave} for a visual representation.
Formally, a cluster $c_i$ undergoes a C\_LEAVE relation if
\begin{itemize}
	\item $c_{i,t-1} \neq \varnothing$
	\item $c_{i,t} \neq \varnothing$
	\item there exists a cluster $c_{j,t}$ such that $|c_{i,t-1} \cap c_{j,t}| \ge min\_cluster$
	\item and $c_{j,t-1} = \varnothing$.
\end{itemize}

The first and the second conditions simply assert that the cluster existed on both timestamps. The third condition states that enough trajectories left the cluster to a form a new cluster. The fourth condition asserts that the new cluster did not exist before. If that was not the case, the relation would be a C\_OUT.

\subsubsection{T\_ENTER}

Intuitively, a cluster $c_i$ undergoes a T\_ENTER relation when a trajectory enters it, either from another cluster or not. See Figure \ref{tenter} for a visual representation.
Formally, a cluster $c_i$ undergoes a T\_ENTER relation if
\begin{itemize}
	\item $traj_{j,t-1} \notin c_{i,t-1}$
	\item $traj_{j,t} \in c_{i,t}$
\end{itemize}
These two conditions are required, but not sufficient. More conditions are required depending on whether the trajectory entered the cluster from another cluster or not. In case $traj_j$ did \textit{not} enter the cluster from another cluster, i.e. $traj_{j,t-1} \notin c_{k,t-1}$ for all $k = 1,\ldots,n$, then one the following two condition is necessary:
\begin{itemize}
	\item $c_{i,t-1} \neq \varnothing$
	\item there exists cluster $c_{k,t-1}$ such that $|c_{i,t} \cap c_{k,t-1}| \ge min\_cluster$.
\end{itemize}

If trajectory $traj_j$ \textit{did} enter the cluster from another cluster, the following two conditions are necessary: let $c_{l,t-1}$ be the cluster $traj_j$ came from, i.e. $traj_{j,t-1} \in c_{l,t-1}$.
\begin{itemize}
	\item $|c_{i,t} \cap c_{l,t-1}| < min\_cluster$
	\item there exists at most one cluster $c_{k,t-1}$ such that $|c_{i,t} \cap c_{k,t-1}| \ge min\_cluster$.
\end{itemize}

The first two conditions simply assert that a trajectory entered a cluster. If this trajectory did not come from a cluster, then the two additional conditions assert that the trajectory is not part of a GROUP relation. If this trajectory did come from a cluster, then the two additional conditions assert that the trajectory is not part of a C\_ENTER (first additional condition) or MERGE (second additional condition) relation.

\subsubsection{T\_LEAVE}

Intuitively, a cluster $c_i$ undergoes a T\_LEAVE relation when one of its trajectories leaves it, either to another cluster or not. See Figure \ref{tleave} for a visual representation.
Formally, a cluster $c_i$ undergoes a T\_LEAVE relation if
\begin{itemize}
	\item $traj_{j,t-1} \in c_{i,t-1}$
	\item $traj_{j,t} \notin c_{i,t}$
\end{itemize}
These two conditions are required, but are not sufficient. More conditions are required depending on whether the trajectory left to another cluster or not. In case $traj_j$ did \textit{not} left to another cluster, i.e. $traj_{j,t} \notin c_{k,t}$ for all $k = 1,\ldots,n$, then one the following two conditions is necessary:
\begin{itemize}
	\item $c_{i,t} \neq \varnothing$
	\item there exists cluster $c_{k,t}$ such that $|c_{i,t-1} \cap c_{k,t}| \ge min\_cluster$.
\end{itemize}

If trajectory $traj_j$ \textit{did} leave to another cluster, the following two conditions are necessary: let $c_{l,t}$ be the cluster $traj_j$ went to, i.e. $traj_{j,t} \in c_{l,t}$.
\begin{itemize}
	\item $|c_{i,t-1} \cap c_{l,t}| < min\_cluster$
	\item there exists at most one cluster $c_{k,t}$ such that $|c_{i,t-1} \cap c_{k,t}| \ge min\_cluster$.
\end{itemize}

The first two conditions simply assert that a trajectory left a cluster. If this trajectory did not leave to a cluster, then the two additional conditions assert that the trajectory is not part of a DISPERSE relation. If this trajectory did leave to a cluster, then the two additional conditions assert that the trajectory is not part of a C\_LEAVE (first additional condition) or SPLIT (second additional condition) relation.

\subsubsection{C\_IN}

Intuitively, a cluster $c_i$ undergoes a C\_IN relation when a group of trajectories (enough to form a new cluster) leaves a cluster and enters $c_i$, immediately after. See Figure \ref{cin} for a visual representation.
Formally, a cluster $c_i$ undergoes a C\_IN relation if
\begin{itemize}
	\item $c_{i,t} \neq \varnothing$
	\item there exists a cluster $c_{j,t-1}$ such that $|c_{i,t} \cap c_{j,t-1}| \ge min\_cluster$
	\item $c_{j,t-1} \neq \varnothing$
\end{itemize}

The first condition simply asserts that the receiving cluster exists after the transition. The second condition states that the number of trajectories exchanged is large enough to be deemed a cluster. The third condition states that the cluster who sent trajectories existed before the transition.

\subsubsection{C\_OUT}

Intuitively, a cluster $c_i$ undergoes a C\_OUT relation when a group of trajectories (enough to form a new cluster) leaves $c_i$ and enters another cluster, immediately after. See Figure \ref{cout} for a visual representation.
Formally, a cluster $c_i$ undergoes a C\_OUT relation if
\begin{itemize}
	\item $c_{i,t-1} \neq \varnothing$
	\item there exists a cluster $c_{j,t}$ such that $|c_{i,t-1} \cap c_{j,t}| \ge min\_cluster$
	\item $c_{j,t} \neq \varnothing$
\end{itemize}

The first condition simply asserts that the cluster sending trajectories exists before the transition. The second condition states that the number of trajectories exchanged is large enough to be deemed a cluster. The third condition states that the receiving cluster exists after the transition.

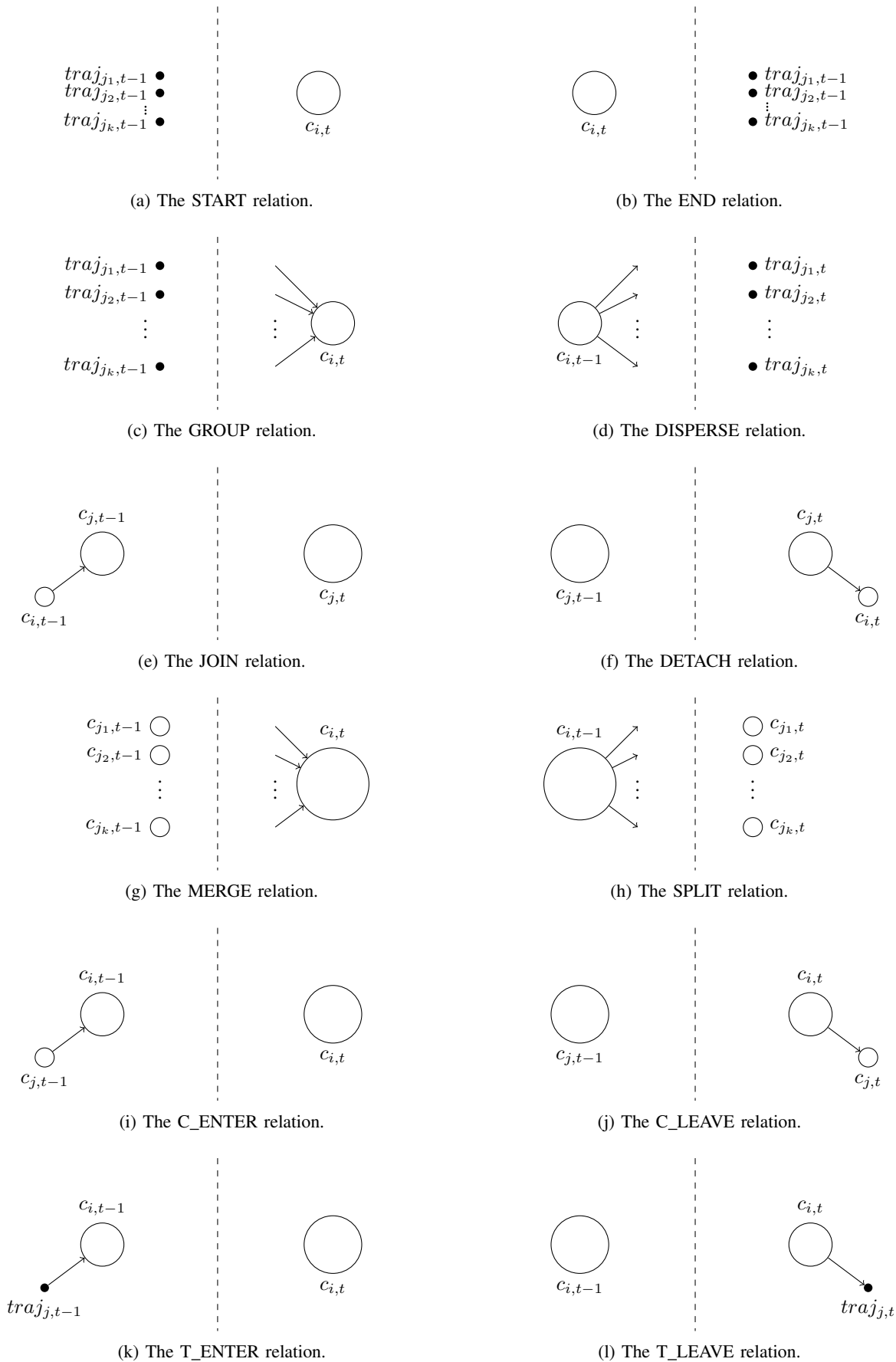
\begin{figure*}
\centering
\begin{subfigure}{0.45\linewidth}
\begin{tikzpicture}
\draw [opacity=0] (-4,1.5) -- (4,1.5) -- (4,-1.5) -- (-4,-1.5) -- (-4,1.5); 

\node (c1) at (1.75,0.0) [draw, circle, minimum size=0.75cm, label=below:$c_{i,t}$] {};

\node (n1) at (-1, 0.30) [circle, fill=black, inner sep=1.5pt, minimum size=2pt, label=left:$traj_{j_1,t-1}$]{};
\node (n2) at (-1, 0.00) [circle, fill=black, inner sep=1.5pt, minimum size=2pt, label=left:$traj_{j_2,t-1}$]{};
\node (n3) at (-1,-0.50) [circle, fill=black, inner sep=1.5pt, minimum size=2pt, label=left:$traj_{j_k,t-1}$]{};

\node (n41) at (-1.25,-0.24) [circle, fill=black, inner sep=0.35pt, minimum size=0.35pt]{};
\node (n42) at (-1.25,-0.30) [circle, fill=black, inner sep=0.35pt, minimum size=0.35pt]{};
\node (n43) at (-1.25,-0.36) [circle, fill=black, inner sep=0.35pt, minimum size=0.35pt]{};

\draw [dashed] (0,1.5) -- (0,-1.5);

\end{tikzpicture}
\caption {The START relation.}
\label{start}
\end{subfigure}
\begin{subfigure}{0.45\linewidth}
\begin{tikzpicture}
\draw [opacity=0] (-4,1.5) -- (4,1.5) -- (4,-1.5) -- (-4,-1.5) -- (-4,1.5); 

\node (c1) at (-1.75,0.0) [draw, circle, minimum size=0.75cm, label=below:$c_{i,t}$] {};

\node (n1) at (1, 0.30) [circle, fill=black, inner sep=1.5pt, minimum size=2pt, label=right:$traj_{j_1,t-1}$]{};
\node (n2) at (1, 0.00) [circle, fill=black, inner sep=1.5pt, minimum size=2pt, label=right:$traj_{j_2,t-1}$]{};
\node (n3) at (1,-0.50) [circle, fill=black, inner sep=1.5pt, minimum size=2pt, label=right:$traj_{j_k,t-1}$]{};

\node (n41) at (1.25,-0.18) [circle, fill=black, inner sep=0.35pt, minimum size=0.35pt]{};
\node (n42) at (1.25,-0.24) [circle, fill=black, inner sep=0.35pt, minimum size=0.35pt]{};
\node (n43) at (1.25,-0.30) [circle, fill=black, inner sep=0.35pt, minimum size=0.35pt]{};

\draw [dashed] (0,1.5) -- (0,-1.5);

\end{tikzpicture}
\caption {The END relation.}
\label{end}
\end{subfigure}

\par\bigskip

\begin{subfigure}{0.45\linewidth}
\begin{tikzpicture}
\draw [opacity=0] (-4,1.5) -- (4,1.5) -- (4,-1.5) -- (-4,-1.5) -- (-4,1.5); 

\node (c1) at (2,0) [draw, circle, minimum size=0.75cm, label=below:$c_{i,t}$] {};

\coordinate (t1) at (1,1.0);
\coordinate (t2) at (1,0.5);
\coordinate (t3) at (1,-0.75);
\node       (t4) at (1.0,0) {$\vdots$};

\draw [->] (t1) -- (c1);
\draw [->] (t2) -- (c1);
\draw [->] (t3) -- (c1);

\node (n1) at (-1,1.0)   [circle, fill=black, inner sep=1.5pt, minimum size=2pt, label=left:$traj_{j_1,t-1}$]{};
\node (n2) at (-1,0.5)   [circle, fill=black, inner sep=1.5pt, minimum size=2pt, label=left:$traj_{j_2,t-1}$]{};
\node (n3) at (-1,-0.75) [circle, fill=black, inner sep=1.5pt, minimum size=2pt, label=left:$traj_{j_k,t-1}$]{};
\node (n4) at (-1.25,0)  {$\vdots$};

\draw [dashed] (0,1.5) -- (0,-1.5);

\end{tikzpicture}
\caption {The GROUP relation.}
\label{group}
\end{subfigure}
\begin{subfigure}{0.45\linewidth}
\begin{tikzpicture}
\draw [opacity=0] (-4,1.5) -- (4,1.5) -- (4,-1.5) -- (-4,-1.5) -- (-4,1.5); 

\node (c1) at (-2,0) [draw, circle, minimum size=0.75cm, label=below:$c_{i,t-1}$] {};

\coordinate (t1) at (-1,1.0);
\coordinate (t2) at (-1,0.5);
\coordinate (t3) at (-1,-0.75);
\node       (t4) at (-1.0,0) {$\vdots$};

\draw [->] (c1) -- (t1);
\draw [->] (c1) -- (t2);
\draw [->] (c1) -- (t3);

\node (n1) at (1,1.0)   [circle, fill=black, inner sep=1.5pt, minimum size=2pt, label=right:$traj_{j_1,t}$]{};
\node (n2) at (1,0.5)   [circle, fill=black, inner sep=1.5pt, minimum size=2pt, label=right:$traj_{j_2,t}$]{};
\node (n3) at (1,-0.75) [circle, fill=black, inner sep=1.5pt, minimum size=2pt, label=right:$traj_{j_k,t}$]{};
\node (n4) at (1,0)     [label=right:$\vdots$] {};

\draw [dashed] (0,1.5) -- (0,-1.5);

\end{tikzpicture}
\caption {The DISPERSE relation.}
\label{disperse}
\end{subfigure}

\par\bigskip

\begin{subfigure}{0.45\linewidth}
\begin{tikzpicture}
\draw [opacity=0] (-4,1.5) -- (4,1.5) -- (4,-1.5) -- (-4,-1.5) -- (-4,1.5); 

\node (c1) at (-2,0)     [draw, circle, minimum size=0.75cm, label=above:$c_{j,t-1}$] {};
\node (c2) at (-3,-0.75) [draw, circle, minimum size=0.25cm, label=below:$c_{i,t-1}$] {};

\draw [->] (c2) -- (c1);

\node (c3) at (2,0)      [draw, circle, minimum size=1cm, label=below:$c_{j,t}$] {};

\draw [dashed] (0,1.5) -- (0,-1.5);

\end{tikzpicture}
\caption {The JOIN relation.}
\label{join}
\end{subfigure}
\begin{subfigure}{0.45\linewidth}
\begin{tikzpicture}
\draw [opacity=0] (-4,1.5) -- (4,1.5) -- (4,-1.5) -- (-4,-1.5) -- (-4,1.5); 

\node (c1) at (2,0)     [draw, circle, minimum size=0.75cm, label=above:$c_{j,t}$] {};
\node (c2) at (3,-0.75) [draw, circle, minimum size=0.25cm, label=below:$c_{i,t}$] {};

\draw [->] (c1) -- (c2);

\node (c3) at (-2,0)    [draw, circle, minimum size=1cm, label=below:$c_{j,t-1}$] {};

\draw [dashed] (0,1.5) -- (0,-1.5);

\end{tikzpicture}
\caption {The DETACH relation.}
\label{detach}
\end{subfigure}

\par\bigskip

\begin{subfigure}{0.45\linewidth}
\begin{tikzpicture}
\draw [opacity=0] (-4,1.5) -- (4,1.5) -- (4,-1.5) -- (-4,-1.5) -- (-4,1.5); 

\node (c1) at (2,0) [draw, circle, minimum size=1.25cm, label=above:$c_{i,t}$] {};

\coordinate (co1) at (1,1.0);
\coordinate (co2) at (1,0.5);
\coordinate (co3) at (1,-0.75);
\node       (co4) at (1,0) {$\vdots$};

\draw [->] (co1) -- (c1);
\draw [->] (co2) -- (c1);
\draw [->] (co3) -- (c1);

\node (cc1) at (-1,1.0)   [draw, circle, minimum size=0.25cm, label=left:$c_{j_1,t-1}$] {};
\node (cc2) at (-1,0.5)   [draw, circle, minimum size=0.25cm, label=left:$c_{j_2,t-1}$] {};
\node (cc3) at (-1,-0.75) [draw, circle, minimum size=0.25cm, label=left:$c_{j_k,t-1}$] {};
\node (cc4) at (-1,0)     {$\vdots$};

\draw [dashed] (0,1.5) -- (0,-1.5);

\end{tikzpicture}
\caption {The MERGE relation.}
\label{merge}
\end{subfigure}
\begin{subfigure}{0.45\linewidth}
\begin{tikzpicture}
\draw [opacity=0] (-4,1.5) -- (4,1.5) -- (4,-1.5) -- (-4,-1.5) -- (-4,1.5); 

\node (c1) at (-2,0) [draw, circle, minimum size=1.25cm, label=above:$c_{i,t-1}$] {};

\coordinate (co1) at (-1,1.0);
\coordinate (co2) at (-1,0.5);
\coordinate (co3) at (-1,-0.75);
\node       (co4) at (-1,0) {$\vdots$};

\draw [->] (c1) -- (co1);
\draw [->] (c1) -- (co2);
\draw [->] (c1) -- (co3);

\node (cc1) at (1,1.0)   [draw, circle, minimum size=0.25cm, label=right:$c_{j_1,t}$] {};
\node (cc2) at (1,0.5)   [draw, circle, minimum size=0.25cm, label=right:$c_{j_2,t}$] {};
\node (cc3) at (1,-0.75) [draw, circle, minimum size=0.25cm, label=right:$c_{j_k,t}$] {};
\node (cc4) at (1,0)     {$\vdots$};

\draw [dashed] (0,1.5) -- (0,-1.5);

\end{tikzpicture}
\caption {The SPLIT relation.}
\label{split}
\end{subfigure}

\par\bigskip

\begin{subfigure}{0.45\linewidth}
\begin{tikzpicture}
\draw [opacity=0] (-4,1.5) -- (4,1.5) -- (4,-1.5) -- (-4,-1.5) -- (-4,1.5); 

\node (c1) at (-2,0)     [draw, circle, minimum size=0.75cm, label=above:$c_{i,t-1}$] {};
\node (c2) at (-3,-0.75) [draw, circle, minimum size=0.25cm, label=below:$c_{j,t-1}$] {};

\draw [->] (c2) -- (c1);

\node (c3) at (2,0)      [draw, circle, minimum size=1cm, label=below:$c_{i,t}$] {};

\draw [dashed] (0,1.5) -- (0,-1.5);

\end{tikzpicture}
\caption {The C\_ENTER relation.}
\label{center}
\end{subfigure}
\begin{subfigure}{0.45\linewidth}
\begin{tikzpicture}
\draw [opacity=0] (-4,1.5) -- (4,1.5) -- (4,-1.5) -- (-4,-1.5) -- (-4,1.5); 

\node (c1) at (2,0)     [draw, circle, minimum size=0.75cm, label=above:$c_{i,t}$] {};
\node (c2) at (3,-0.75) [draw, circle, minimum size=0.25cm, label=below:$c_{j,t}$] {};

\draw [->] (c1) -- (c2);

\node (c3) at (-2,0)    [draw, circle, minimum size=1cm, label=below:$c_{j,t-1}$] {};

\draw [dashed] (0,1.5) -- (0,-1.5);

\end{tikzpicture}
\caption {The C\_LEAVE relation.}
\label{cleave}
\end{subfigure}

\par\bigskip

\begin{subfigure}{0.45\linewidth}
\begin{tikzpicture}
\draw [opacity=0] (-4,1.5) -- (4,1.5) -- (4,-1.5) -- (-4,-1.5) -- (-4,1.5); 

\node (c1) at (-2,0)     [draw, circle, minimum size=0.75cm, label=above:$c_{i,t-1}$] {};
\node (t1) at (-3,-0.75) [circle, fill=black, inner sep=1.5pt, minimum size=2pt, label=below:$traj_{j,t-1}$]{};

\draw [->] (t1) -- (c1);

\node (c2) at (2,0)      [draw, circle, minimum size=1cm, label=below:$c_{i,t}$] {};

\draw [dashed] (0,1.5) -- (0,-1.5);

\end{tikzpicture}
\caption {The T\_ENTER relation.}
\label{tenter}
\end{subfigure}
\begin{subfigure}{0.45\linewidth}
\begin{tikzpicture}
\draw [opacity=0] (-4,1.5) -- (4,1.5) -- (4,-1.5) -- (-4,-1.5) -- (-4,1.5); 

\node (c1) at (2,0)     [draw, circle, minimum size=0.75cm, label=above:$c_{i,t}$] {};
\node (t1) at (3,-0.75) [circle, fill=black, inner sep=1.5pt, minimum size=2pt, label=below:$traj_{j,t}$]{};

\draw [->] (c1) -- (t1);

\node (c2) at (-2,0)    [draw, circle, minimum size=1cm, label=below:$c_{i,t-1}$] {};

\draw [dashed] (0,1.5) -- (0,-1.5);

\end{tikzpicture}
\caption {The T\_LEAVE relation.}
\label{tleave}
\end{subfigure}
\caption {The spatial-temporal relations.}
\label{fig:relations1}
\end{figure*}

\begin{figure*}\ContinuedFloat
\centering
\begin{subfigure}{0.45\linewidth}
\begin{tikzpicture}
\draw [opacity=0] (-4,1.5) -- (4,1.5) -- (4,-1.5) -- (-4,-1.5) -- (-4,1.5); 

\node (c1) at (-2,-0.75) [draw, circle, minimum size=0.75cm, label=below:$c_{j,t-1}$] {};
\node (c2) at (-2, 0.75) [draw, circle, minimum size=0.50cm, label=above:$c_{i,t-1}$] {};

\draw [->] (c1) -- (c2);

\node (cc1) at (2,-0.75) [draw, circle, minimum size=0.50cm, label=below:$c_{j,t}$] {};
\node (cc2) at (2, 0.75) [draw, circle, minimum size=0.75cm, label=above:$c_{i,t}$] {};

\draw [dashed] (0,1.5) -- (0,-1.5);

\end{tikzpicture}
\caption {The C\_IN relation.}
\label{cin}
\end{subfigure}
\begin{subfigure}{0.45\linewidth}
\begin{tikzpicture}
\draw [opacity=0] (-4,1.5) -- (4,1.5) -- (4,-1.5) -- (-4,-1.5) -- (-4,1.5); 

\node (c1) at (-2,-0.75) [draw, circle, minimum size=0.75cm, label=below:$c_{i,t-1}$] {};
\node (c2) at (-2, 0.75) [draw, circle, minimum size=0.50cm, label=above:$c_{j,t-1}$] {};

\draw [->] (c1) -- (c2);

\node (cc1) at (2,-0.75) [draw, circle, minimum size=0.50cm, label=below:$c_{i,t}$] {};
\node (cc2) at (2, 0.75) [draw, circle, minimum size=0.75cm, label=above:$c_{j,t}$] {};

\draw [dashed] (0,1.5) -- (0,-1.5);

\end{tikzpicture}
\caption {The C\_OUT relation.}
\label{cout}
\end{subfigure}
\caption {The spatial-temporal relations. \textit{Continued from previous page.}}
\label{fig:relations2}
\end{figure*}
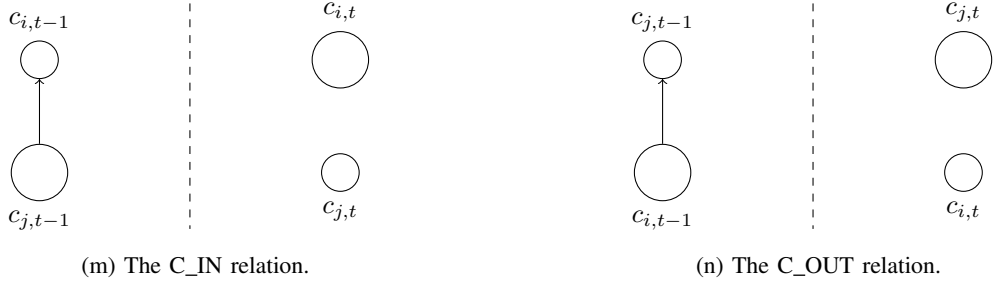

\section{Examples}
\label{examples}

We now illustrate the applicability of these definitions with two examples that have ambiguous interpretations. In the examples, we use a function, derived from the Jaccard index \cite{Jaccard1901547}, to calculate the similarity between clusters in different timestamps. Let $c-i$ and $c_j$ be two clusters in different consecutive timestamps (we omit the timestmap from the notation for clarity). These clusters are the same if
\[ \frac{|c_i \cap c_j|}{|c_i^*|} \ge min\_shared \]
and
\[ \frac{|c_i \cap c_j|}{|c_j^*|} \ge min\_shared \]
where $c_i^*$ and $c_j^*$ are clusters $c_i$ and $c_j$ without trajectories that just entered of left the cluster, $min\_shared$ is the minimum ratio of shared elements (e.g. 0.8 for 80\%), and $|\cdot|$ is an operator to count the number of elements of a cluster.

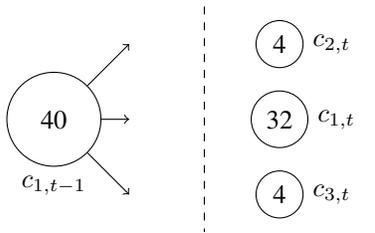
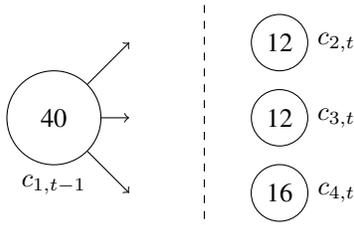
\begin{figure}
\centering
\begin{subfigure}{\linewidth}
\begin{tikzpicture}
\draw [opacity=0] (-4,1.5) -- (4,1.5) -- (4,-1.5) -- (-4,-1.5) -- (-4,1.5); 

\node (c1) at (-2,0) [draw, circle, minimum size=1.25cm, label=below:$c_{1,t-1}$] {40};
\coordinate (co1) at (-1, 1);
\coordinate (co2) at (-1, 0);
\coordinate (co3) at (-1,-1);

\node (c2) at (1, 1) [draw, circle, minimum size=0.5cm, label=right:$c_{2,t}$] {4};
\node (c3) at (1, 0) [draw, circle, minimum size=0.5cm, label=right:$c_{1,t}$] {32};
\node (c4) at (1,-1) [draw, circle, minimum size=0.5cm, label=right:$c_{3,t}$] {4};

\draw[->] (c1) -- (co1);
\draw[->] (c1) -- (co2);
\draw[->] (c1) -- (co3);

\draw [dashed] (0,1.5) -- (0,-1.5);
\end{tikzpicture}
\caption{Two C\_LEAVE relations}
\label{example1a}
\end{subfigure}
\par\bigskip
\begin{subfigure}{\linewidth}
\begin{tikzpicture}
\draw [opacity=0] (-4,1.5) -- (4,1.5) -- (4,-1.5) -- (-4,-1.5) -- (-4,1.5); 

\node (c1) at (-2,0) [draw, circle, minimum size=1.25cm, label=below:$c_{1,t-1}$] {40};
\coordinate (co1) at (-1, 1);
\coordinate (co2) at (-1, 0);
\coordinate (co3) at (-1,-1);

\node (c2) at (1, 1) [draw, circle, minimum size=0.5cm, label=right:$c_{2,t}$] {12};
\node (c3) at (1, 0) [draw, circle, minimum size=0.5cm, label=right:$c_{3,t}$] {12};
\node (c4) at (1,-1) [draw, circle, minimum size=0.5cm, label=right:$c_{4,t}$] {16};

\draw[->] (c1) -- (co1);
\draw[->] (c1) -- (co2);
\draw[->] (c1) -- (co3);

\draw [dashed] (0,1.5) -- (0,-1.5);
\end{tikzpicture}
\caption{A SPLIT relation}
\label{example1b}
\end{subfigure}
\caption {Example 1.}
\label{example1}
\end{figure}

The first example relates to the situation shown in Figure \ref{example1}. During timestamp $t-1$, a cluster $c_1$ contains 40 trajectories. The cluster is divided into three other clusters during timestamp $t$. We assume that a valid cluster has at least three trajectories, i.e. $min\_cluster = 3$. Our intuition seems to resolve each case (the top and the bottom situations) reasonably well, and the notation under each cluster gives away what happened. An automated approach for recognizing what happened should follow our intuition. Using  the relations described in this study, we first start by analyzing the situation in Figure \ref{example1a}. The first step is calculate cross-temporal cluster identity. We assume $min\_shared = 0.8$. Since
\[\frac{|c_{1,t-1} \cap c_{1,t}|}{|c_{1,t-1}|} = \frac{32}{40} = 0.8 \ge 0.8 = min\_shared \]
\[\frac{|c_{1,t-1} \cap c_{1,t}|}{|c_{1,t}|}   = \frac{32}{32} = 1   \ge 0.8 = min\_shared \]
we, as expected, identify that $c_{1,t-1}$ and $c_{1,t}$ are the same cluster. Note that the cross temporal cluster identity test fails for $c_{2,t}$ and $c_{3,t}$, as expected. Intuitively, we understand that $c_{2,t}$ and $c_{3,t}$ should be regarded as clusters that left $c_{1,t-1}$. Based on the definition of the C\_LEAVE relation with $i = 1$, and $j = 2$ (the case $j = 3$ is analogous),

\begin{itemize}
	\item $c_{1,t-1} \neq \varnothing$
	\item $c_{1,t}   \neq \varnothing$
	\item There exists a cluster $c_{2,t}$ such that $|c_{1,t-1} \cap c_{2,t}| \ge min\_cluster$
	\item $c_{2,t-1} = \varnothing$
\end{itemize}

In the situation of Figure \ref{example1b}, the cross-temporal cluster identity test fails for all three clusters $c_{1,t}$, $c_{2,t}$, and $c_{3,t}$. A SPLIT relation is underway, confirmed by the following tests from the SPLIT definition, where $i = 1$ and $j = 2$ and $k = 3$

\begin{itemize}
	\item $c_{i,t-1} \neq \varnothing$
	\item $c_{i,t} = \varnothing$
	\item There exists at least two clusters $c_{2,t}$ and $c_{3,t}$ such that $|c_{1,t-1} \cap c_{2,t}| \ge min\_cluster$ and $|c_{1,t-1} \cap c_{3,t}| \ge min\_cluster$
\end{itemize}

For each situation, other types of relation do not apply to $c_1$. Also, note that other relations do apply to the other clusters in each situation. For example, $c_{2,t}$ in Figure \ref{example1b} undergoes a DETACH relation as noted below, following the definition of the relation with $i=2$ and $j=1$

\begin{itemize}
	\item $c_{2,t-1} = \varnothing$
	\item $c_{2,t} \neq \varnothing$
	\item There exists only one cluster $c_{1,t-1}$ such that $|c_{2,t} \cap c_{1,t-1}| \ge min\_cluster$
\end{itemize}

\begin{figure}
\centering
\begin{subfigure}{\linewidth}
\begin{tikzpicture}
\draw [opacity=0] (-4,2) -- (4,2) -- (4,-2) -- (-4,-2) -- (-4,2); 

\node (c1) at (-2,0) [draw, circle, minimum size=1.25cm, label=below:$c_{1,t-1}$] {20};
\node (c2) at ( 2,1) [draw, circle, minimum size=0.5cm,  label=below:$c_{2,t}$  ] { 7};
\node (c3) at (2,-1) [draw, circle, minimum size=0.5cm,  label=below:$c_{1,t}$  ] {16};

\node (t1) at (-2,1.5) [circle, fill=black, inner sep=1.5pt, minimum size=2pt, label=left:$traj_{1,t-1}$]{};
\node (t2) at (-2,1.3) [circle, fill=black, inner sep=1.5pt, minimum size=2pt, label=left:$traj_{2,t-1}$]{};
\node (t3) at (-2,1.1) [circle, fill=black, inner sep=1.5pt, minimum size=2pt, label=left:$traj_{3,t-1}$]{};

\coordinate (co1) at (1, 1.5);
\coordinate (co2) at (1, 1.3);
\coordinate (co3) at (1, 1.1);

\draw[->] (co1) -- (c2);
\draw[->] (co2) -- (c2);
\draw[->] (co3) -- (c2);

\coordinate (co4) at (-1, 0.3);
\coordinate (co5) at (-1,-0.3);

\draw [->] (c1) -- (co4);
\draw [->] (c1) -- (co5);

\draw [dashed] (0,1.5) -- (0,-1.5);
\end{tikzpicture}
\caption{A C\_LEAVE and a T\_ENTER relation.}
\label{example2a}
\end{subfigure}
\par\bigskip
\begin{subfigure}{\linewidth}
\begin{tikzpicture}
\draw [opacity=0] (-4,2) -- (4,2) -- (4,-2) -- (-4,-2) -- (-4,2); 

\node (c1) at (-2,0) [draw, circle, minimum size=1.25cm, label=below:$c_{1,t-1}$] {20};
\node (c2) at ( 2,1) [draw, circle, minimum size=0.5cm,  label=below:$c_{2,t}$  ] {13};
\node (c3) at (2,-1) [draw, circle, minimum size=0.5cm,  label=below:$c_{3,t}$  ] {10};

\node (t1) at (-2,1.5) [circle, fill=black, inner sep=1.5pt, minimum size=2pt, label=left:$traj_{1,t-1}$]{};
\node (t2) at (-2,1.3) [circle, fill=black, inner sep=1.5pt, minimum size=2pt, label=left:$traj_{2,t-1}$]{};
\node (t3) at (-2,1.1) [circle, fill=black, inner sep=1.5pt, minimum size=2pt, label=left:$traj_{3,t-1}$]{};

\coordinate (co1) at (1, 1.5);
\coordinate (co2) at (1, 1.3);
\coordinate (co3) at (1, 1.1);

\draw[->] (co1) -- (c2);
\draw[->] (co2) -- (c2);
\draw[->] (co3) -- (c2);

\coordinate (co4) at (-1, 0.3);
\coordinate (co5) at (-1,-0.3);

\draw [->] (c1) -- (co4);
\draw [->] (c1) -- (co5);

\draw [dashed] (0,1.5) -- (0,-1.5);
\end{tikzpicture}
\caption{A SPLIT and a T\_ENTER relation.}
\label{example2b}
\end{subfigure}
\par\bigskip
\begin{subfigure}{\linewidth}
\begin{tikzpicture}
\draw [opacity=0] (-4,2) -- (4,2) -- (4,-2) -- (-4,-2) -- (-4,2); 

\node (c1) at (-2,0) [draw, circle, minimum size=1.25cm, label=below:$c_{1,t-1}$] {20};
\node (c2) at ( 2,1) [draw, circle, minimum size=0.5cm,  label=below:$c_{2,t}$  ] { 4};
\node (c3) at (2,-1) [draw, circle, minimum size=0.5cm,  label=below:$c_{1,t}$  ] {19};

\node (t1) at (-2,1.5) [circle, fill=black, inner sep=1.5pt, minimum size=2pt, label=left:$traj_{1,t-1}$]{};
\node (t2) at (-2,1.3) [circle, fill=black, inner sep=1.5pt, minimum size=2pt, label=left:$traj_{2,t-1}$]{};
\node (t3) at (-2,1.1) [circle, fill=black, inner sep=1.5pt, minimum size=2pt, label=left:$traj_{3,t-1}$]{};

\coordinate (co1) at (1, 1.5);
\coordinate (co2) at (1, 1.3);
\coordinate (co3) at (1, 1.1);

\draw[->] (co1) -- (c2);
\draw[->] (co2) -- (c2);
\draw[->] (co3) -- (c2);

\coordinate (co4) at (-1, 0.3);
\coordinate (co5) at (-1,-0.3);

\draw [->] (c1) -- (co4);
\draw [->] (c1) -- (co5);

\draw [dashed] (0,1.5) -- (0,-1.5);
\end{tikzpicture}
\caption{A T\_LEAVE and a GROUP relation.}
\label{example2c}
\end{subfigure}
\caption{Example 2.}
\label{example2}
\end{figure}
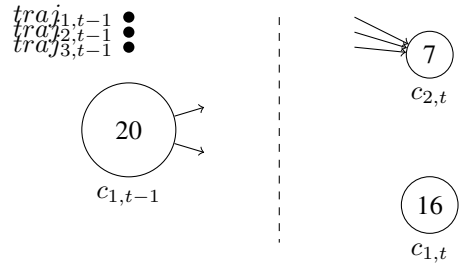
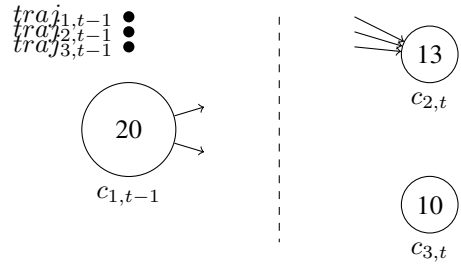
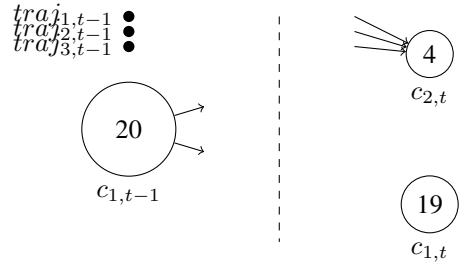

The second, more complex, example is illustrated in Figure \ref{example2}. In each situation, there are a cluster and three trajectories at timestamp $t-1$, which, after some relations, become just two clusters at timestamp $t$. We are mainly interested in describing what relation happened to $c_{1,t-1}$ and to $c_{2,t}$. We assume $min\_cluster=3$ and $min\_shared=0.8$. Again, we start by calculating the cross-temporal cluster identity for $c_{1,t-1}$ in each case. The calculations are not printed here, but the notation near each cluster shows the results. Consider Figure \ref{example2a}. After identifying $c_1$ across timestamps, we notice the cluster underwent a C\_LEAVE relation, creating $c_2$. We also notice that $c_{1,t}$ is not a new cluster, we perceive the three trajectories at timestamp $t-1$ as entering $c_{2,t}$, causing it to undergo three T\_ENTER relations. Below are the requirements for a T\_ENTER relation with $i=2$ and $j=1$ (the cases $j=2$ and $j=3$ are analogous).

\begin{itemize}
	\item $traj_{1,t-1} \notin c_{2,t-1}$
	\item $traj_{1,t} \in c_{2,t}$
	\item $c_{2,t-1} \neq \varnothing$
\end{itemize}

Now, consider Figure \ref{example2b}. The cluster $c_1$ does not exist at the timestamp $t$, indicating that $c_{1,t-1}$ underwent a SPLIT relation. In that case, cluster $c_{2,t}$ also underwent three T\_ENTER relations, as explained below with $i=2$, $j=1$, and $k=1$ (the cases $j=2$ and $j=3$ are analogous).

\begin{itemize}
	\item $traj_{1,t-1} \notin c_{2,t-1}$
	\item $traj_{1,t} \in c_{2,t}$
	\item There exists cluster $c_{1,t-1}$ such that $|c_{2,t} \cap c_{1,t-1}| \ge min\_cluster$
\end{itemize}

Finally, in Figure \ref{example2c}, cluster $c_{1,t-1}$ undergoes a T\_LEAVE relation, since it continues as the cluster $c_{1,t}$ with 19 elements. Therefore, cluster $c_{2,t}$ is not formed from cluster $c_{1,t-1}$, as in the previous examples, but from the grouping of many trajectories. Below is the definition of a GROUP relation with $i=2$.

\begin{itemize}
	\item $c_{2,t-1} = \varnothing$
	\item $c_{2,t} \neq \varnothing$
	\item There does not exist cluster $c_{j,t-1}$ such that $|c_{2,t} \cap c_{j,t-1}| \ge min\_cluster$
\end{itemize}

The two previously described examples illustrate the applicability of the spatial-temporal cluster relation definitions. The definitions could distinguish ambiguous cases and produce an intuitive result. In more complex situations, such as when many trajectories are constantly moving between clusters, human intuition may not helpful or feasible, and an automated support is required.

\section{Conclusions}
\label{conclusions}

Spatial-temporal data is highly available and their analysis is of significant importance. One way to analyze this type of data is using clustering approaches. Traditionally, clusters are static and very useful for classification and anomaly detection. More recently, the dynamic nature of clusters, those who move in space across time are being considered in analysis. Unfortunately, traditional and recent clustering approaches limit themselves in the analysis of static clusters. Even the clustering approaches that take into consideration a more dynamic nature of clusters are limited in a sense that only the identification of a dynamic cluster is performed. There exists a lack of approaches that observe clusters from different perspectives.

Spatial-temporal, moving, or dynamic clusters can be described in terms of their relations with other elements, that is clusters or elements contained in the cluster. These relations summarize important information about the lifetime of a cluster and leads to novel perspectives on cluster dynamics and to significant research opportunities.

This report introduces, describes, and formalizes 14 spatial-temporal cluster relations that may happened during the existence of a cluster. Two of these relations are existential and do not involve other elements. The remaining 12 relations describe interactions between two spatial-temporal clusters or a spatial-temporal cluster and one of its elements. Here, one cluster element is assumed to be a trajectory, that is, a spatial-temporal point that keeps their entire history of movement. A spatial-temporal cluster is a set of trajectories that are similar, according to the clustering approach. For example, for a density-based clustering approach, the distance between elements and the number of elements within a radius are crucial for the definition of a cluster. We concluded our description and formalization with two examples designed to be ambiguous that can happen in any domain. The set of relations is able to identify each situation and describe them in a way that matches our intuition.

Finally, we provide some avenues for further research based on the new spatial-temporal relations. Although the set of proposed relations do assist in identifying events that occur during the existence of a spatial-temporal cluster, this set is not argued to be complete and this work can be extended in several ways. First, novel approaches could be provided to support automated cluster analysis based on the relations. We have conducted preliminary work in this direction. Second, novel spatial-temporal relations can be further derived to capture more complex situations. The proposed relations are limited to a two-dimensional world in which moving objects are considered points navigating in a 2D space. Novel relations can be created when higher dimensions are taken into account, for instance, when 3D moving objects navigate in a 3D space. Finally, it would be interesting to investigate how the proposed relations can be adapted to cases outside of the geographical context in domains such as microbiology or astronomy.

\section*{Acknowledgment}
The authors thank the Natural Sciences and Engineering Research Council of Canada (NSERC), and the Ontario Research Fund of the Ontario Ministry of Research, Innovation, and Science for their financial support for this research.

\balance

\bibliographystyle{IEEEtran}
\bibliography{./scopus,./noscopus}

\begin{thebibliography}{10}
\providecommand{\url}[1]{#1}
\csname url@samestyle\endcsname
\providecommand{\newblock}{\relax}
\providecommand{\bibinfo}[2]{#2}
\providecommand{\BIBentrySTDinterwordspacing}{\spaceskip=0pt\relax}
\providecommand{\BIBentryALTinterwordstretchfactor}{4}
\providecommand{\BIBentryALTinterwordspacing}{\spaceskip=\fontdimen2\font plus
\BIBentryALTinterwordstretchfactor\fontdimen3\font minus
  \fontdimen4\font\relax}
\providecommand{\BIBforeignlanguage}[2]{{%
\expandafter\ifx\csname l@#1\endcsname\relax
\typeout{** WARNING: IEEEtran.bst: No hyphenation pattern has been}%
\typeout{** loaded for the language `#1'. Using the pattern for}%
\typeout{** the default language instead.}%
\else
\language=\csname l@#1\endcsname
\fi
#2}}
\providecommand{\BIBdecl}{\relax}
\BIBdecl

\bibitem{Wang2000202}
X.~Wang, X.~Zhou, and S.~Lu, ``Spatiotemporal data modelling and management: A
  survey,'' \emph{Proceedings of the Conference on Technology of
  Object-Oriented Languages and Systems, TOOLS}, pp. 202--211, 2000.

\bibitem{Kent2002213}
J.~Kent and K.~Mardia, \emph{Modelling strategies for spatial-temporal
  data}.\hskip 1em plus 0.5em minus 0.4em\relax CRC Press, 2002.

\bibitem{Tan2018}
P.-N. Tan, M.~Steinbach, A.~Karpatne, and V.~Kumar, \emph{Introduction to Data
  Mining}.\hskip 1em plus 0.5em minus 0.4em\relax Pearson, 2018.

\bibitem{Lloyd1982129}
S.~Lloyd, ``Least squares quantization in pcm,'' \emph{IEEE Transactions on
  Information Theory}, vol.~28, no.~2, pp. 129--137, 1982.

\bibitem{Ester1996}
M.~Ester, H.-P. Kriegel, J.~Sander, and X.~Xu, ``A density-based algorithm for
  discovering clusters in large spatial databases with noise,'' in
  \emph{Proceedings of the second international conference on knowledge
  discovery and data mining}, 1996, pp. 226--231.

\bibitem{Kisilevich2010}
S.~Kisilevich, F.~Mansmann, M.~Nanni, and S.~Rinzivillo, \emph{Spatio-temporal
  clustering}.\hskip 1em plus 0.5em minus 0.4em\relax Springer US, 2010, pp.
  855--874.

\bibitem{Ansari2019}
M.~Ansari, A.~Ahmad, S.~Khan, G.~Bhushan, and Mainuddin, ``Spatiotemporal
  clustering: a review,'' \emph{Artificial Intelligence Review}, 2019.

\bibitem{Birant2007208}
D.~Birant and A.~Kut, ``St-dbscan: An algorithm for clustering spatial-temporal
  data,'' \emph{Data and Knowledge Engineering}, vol.~60, no.~1, pp. 208--221,
  2007.

\bibitem{Chen20152575}
X.~Chen, J.~Faghmous, A.~Khandelwal, and V.~Kumar, ``Clustering dynamic
  spatio-temporal patterns in the presence of noise and missing data,'' in
  \emph{IJCAI International Joint Conference on Artificial Intelligence}, vol.
  2015-January.\hskip 1em plus 0.5em minus 0.4em\relax International Joint
  Conferences on Artificial Intelligence, 2015, pp. 2575--2581.

\bibitem{Forlizzi1998332}
L.~Forlizzi and E.~Nardelli, ``Some results on the modelling of spatial data,''
  \emph{Lecture Notes in Computer Science (including subseries Lecture Notes in
  Artificial Intelligence and Lecture Notes in Bioinformatics)}, vol. 1521, pp.
  332--343, 1998.

\bibitem{Abdul-Rahman20081}
A.~Abdul-Rahman and M.~Pilouk, \emph{Spatial data modelling for 3D GIS}.\hskip
  1em plus 0.5em minus 0.4em\relax Springer Berlin Heidelberg, 2008.

\bibitem{Jensen199936}
C.~Jensen and R.~Snodgrass, ``Temporal data management,'' \emph{IEEE
  Transactions on Knowledge and Data Engineering}, vol.~11, no.~1, pp. 36--44,
  1999.

\bibitem{Laxman2006173}
S.~Laxman and P.~Sastry, ``A survey of temporal data mining,'' \emph{Sadhana -
  Academy Proceedings in Engineering Sciences}, vol.~31, no.~2, pp. 173--198,
  2006.

\bibitem{Clementini1993277}
E.~Clementini, P.~Di~Felice, and P.~Van~Oosterom, ``A small set of formal
  topological relationships suitable for end-user interaction,'' \emph{Lecture
  Notes in Computer Science (including subseries Lecture Notes in Artificial
  Intelligence and Lecture Notes in Bioinformatics)}, vol. 692 LNCS, pp.
  277--295, 1993.

\bibitem{Mazimpaka2017}
J.~D. Mazimpaka, ``A framework for analysing trajectories of movement in a
  dynamic geographic context,'' Ph.D. dissertation, University of Augsburg,
  2017.

\bibitem{Aiello2007}
M.~Aiello, I.~Pratt-Hartmann, and J.~van Benthem, \emph{Handbook of Spatial
  Logistics}.\hskip 1em plus 0.5em minus 0.4em\relax Springer, 2007.

\bibitem{Portugal20193452}
I.~Portugal, P.~Alencar, and D.~Cowan, ``Trajectory cluster lifecycle analysis:
  An evolutionary perspective,'' in \emph{Proceedings - 2018 IEEE International
  Conference on Big Data, Big Data 2018}.\hskip 1em plus 0.5em minus
  0.4em\relax Institute of Electrical and Electronics Engineers Inc., 2019, pp.
  3452--3455.

\bibitem{Portugal20194534}
------, ``A software framework for cluster lifecycle analysis in
  transportation,'' in \emph{Proceedings - 2018 IEEE International Conference
  on Big Data, Big Data 2018}.\hskip 1em plus 0.5em minus 0.4em\relax Institute
  of Electrical and Electronics Engineers Inc., 2019, pp. 4534--4539.

\bibitem{Jaccard1901547}
P.~Jaccard, ``Etude comparative de la distribution florale dans une portion des
  alpes et des jura,'' \emph{Bulletin de la Soci{\'e}t{\'e} Vaudoise des
  Sciences Natturelles}, vol.~37, pp. 547--579, 1901.

\end{thebibliography}

\end{document}